\documentclass{vow2008}
\title{Local Benchmarks of Infrared Galaxy Evolution : the SWIRE-SDSS Database, Far-Infrared Local Luminosity Functions \& Virtual Observatory Tools}
\author[1]{Mattia Vaccari}
\author[1,2]{Lucia Marchetti}
\author[1]{Alberto Franceschini}
\author[2]{Ismael Perez-Fournon}
\affil[1]{Department of Astronomy, University of Padova, Italy, \textbf{mattia@mattiavaccari.net}}
\affil[2]{Instituto de Astrofisica de Canarias, La Laguna, Tenerife, Spain}
\usepackage{graphicx}
\usepackage{epstopdf}
\begin{document}


\maketitle

\begin{abstract}
We describe the construction and the properties of the SWIRE-SDSS database, a preliminary derivation of the Far-Infrared Local Luminosity Functions at 24/70/160 micron based on such a database and ways in which VO tools will allow to refine and extend such work.
%
%
\end{abstract}
\section{Infrared Galaxy Surveys Across Cosmic Time}
In the context of ever deeper surveys at most wavelengths, it is becoming increasingly difficult and important to reliably measure galaxy properties in the Local Universe: difficult because the very possibility to carry out extremely deep observations leads to most observing time being spent on the deepest pencil-beam surveys rather than on shallower wider-area ones which are gradually becoming the domain of dedicated facilities (see e.g. 2MASS, SDSS, Pan-STARRS and LSST) and important because the increasingly detailed knowledge of the high-redshift Universe needs similarly well-defined local benchmarks to trace the formation and evolution of galaxies across cosmic time in great detail. Perhaps more importantly, in the era of multi-wavelength surveys and virtual observatories, shallow wide-area surveys with large data rates are likely to profit the most from the paradigm shift caused in astronomical research by the easy access to a number of otherwise separate databases for science exploitation.

This is particularly relevant for infrared galaxy surveys because, as first demonstrated by IRAS, infrared galaxies undergo a dramatic evolution in the Local Universe. As later shown by ISO and Spitzer, such an evolution is sustained up to $z\sim1$, when it subsides leading to an approximately constant infrared luminosity density up to $z\sim3$. Accurately pinpointing the infrared local luminosity function and studying its evolution over the $0<z<1$ range is therefore of the uttermost importance for studies of galaxy formation and evolution.

This work capitalizes on the above trends by exploiting (mainly) two major survey projects, in the infrared and optical respectively, which have recently got close to completion. The SWIRE (infrared) and the SDSS (optical) datasets in the Lockman Hole region, along with ancillary datasets at optical and near-infrared wavelengths such as 2MASS and UKIDSS, are used to derive the galaxy local luminosity function at 24, 70 and 160 $\mu$m and thus place stronger constraints on models for the formation and evolution of infrared galaxies.
\section{The SWIRE-SDSS Database}
The SWIRE (Lonsdale et al. 2003) and SDSS (York et al. 2002) projects provide remarkably well-matched datasets for studies of optical vs infrared galaxy emission. This is particularly true in the Local Universe, where redshifts were measured spectroscopically by the SDSS for the majority of infrared- (as well as optically-) bright sources.

The size of the catalogs produced by these projects at multiple wavelengths, however, requires robust automated procedures to deal with the assessment of catalog quality (and most importantly its reliability and completeness), the association of sources detected at different wavelengths and the optimal handling of multiwavelength photometric and spectroscopic datasets, most of which coming with their own value-added data products such as photometric and spectroscopic redshifts, stellar mass and star formation rate estimates etc.


For our purposes, i.e.\ in order to determine the LLF of FIR-selected, there are two critically important steps while handling the data. Namely, one first needs to determine the correct optical and NIR counterpart of each FIR source and then go on to determine a reliable redshift for each source based on either photometric or spectroscopic means.

The conceptually simple issue of identifying a source in images taken at different times, spatial resolutions or wavelengths, has recently been receiving a great deal of attention, largely due to the increasing source densities provided by modern detectors and coordinated surveys carried out at a number of wavelengths. The nearest-neighbor traditional approach has thus been gradually superseded by the likelihood ratio technique by Sutherland \& Saunders 1992 as well as by other more sophisticated methods such as the fully Bayesian approach by Budavari \& Szalay 2008. Similarly, imaging in 5 or more bands and/or at NIR wavelengths greatly improve the accuracy of photometric redshifts, as demonstrated, between others, by Rowan-Robinson et al. 2008 and Ilbert e al. 2009.

In our case, limiting the study to bright ($S_{24,70,160} > 1,15,75$ mJy) and local ($0<z<0.25$) FIR sources greatly simplified both issues. On one hand, the source density of bright FIR sources is small enough to provide a straightforward optical identification, and particularly so when suitably deep imaging at intermediate IRAC wavelengths are available. On the other hand, spectroscopic redshifts are provided by the SDSS for a large portion of bright local galaxies, while carefully calibrated and validated photometric redshifts are available for virtually all bright sources via the SDSS collaboration.

In this work we therefore used the upcoming SWIRE final data release and the SDSS DR6 in the Lockman field, where the joint coverage by SWIRE and SDSS amounts to $\sim 10$ deg$^2$. Nearest-neighbor search between MIPS70 \& MIPS160 micron FIR single-band catalogs and IRAC+MIPS24 MIR 4-band catalogs guaranteed a reliable and complete association process. Similarly, association between IRAC+MIPS24 MIR catalogs and SDSS 5-band catalogs provided a robust link to the optical, along with a vast spectroscopic and photometric redshift database. A number of ancillary optical (SWIRE team proprietary imaging) and NIR catalogs (2MASS PSC and UKIDSS DR4) were also positionally correlated against SDSS, providing additional source diagnostic power over the full field (albeit with 2MASS PSC only) but also deeper photometry over portions of the field. Source and redshift reliability flags were extracted from the relevant catalogs and included in the database. Stars (only a few at the bright FIR flux limits we were considering) were identified on the basis of a number of criteria, so as to provide an extragalactic source catalog of the uttermost quality.

\begin{figure}[!h]
\centering
\includegraphics[width=\linewidth]{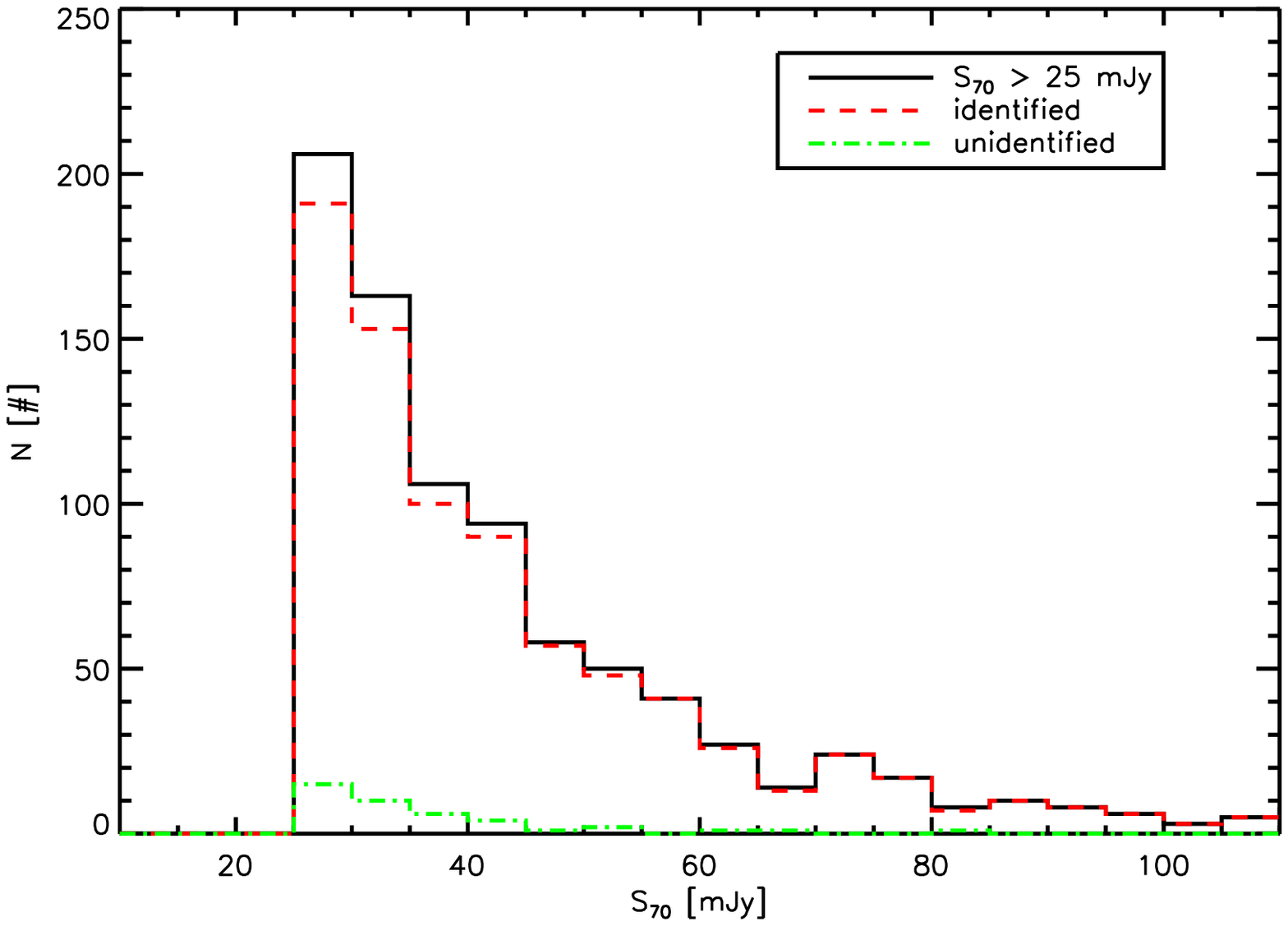}
\includegraphics[width=\linewidth]{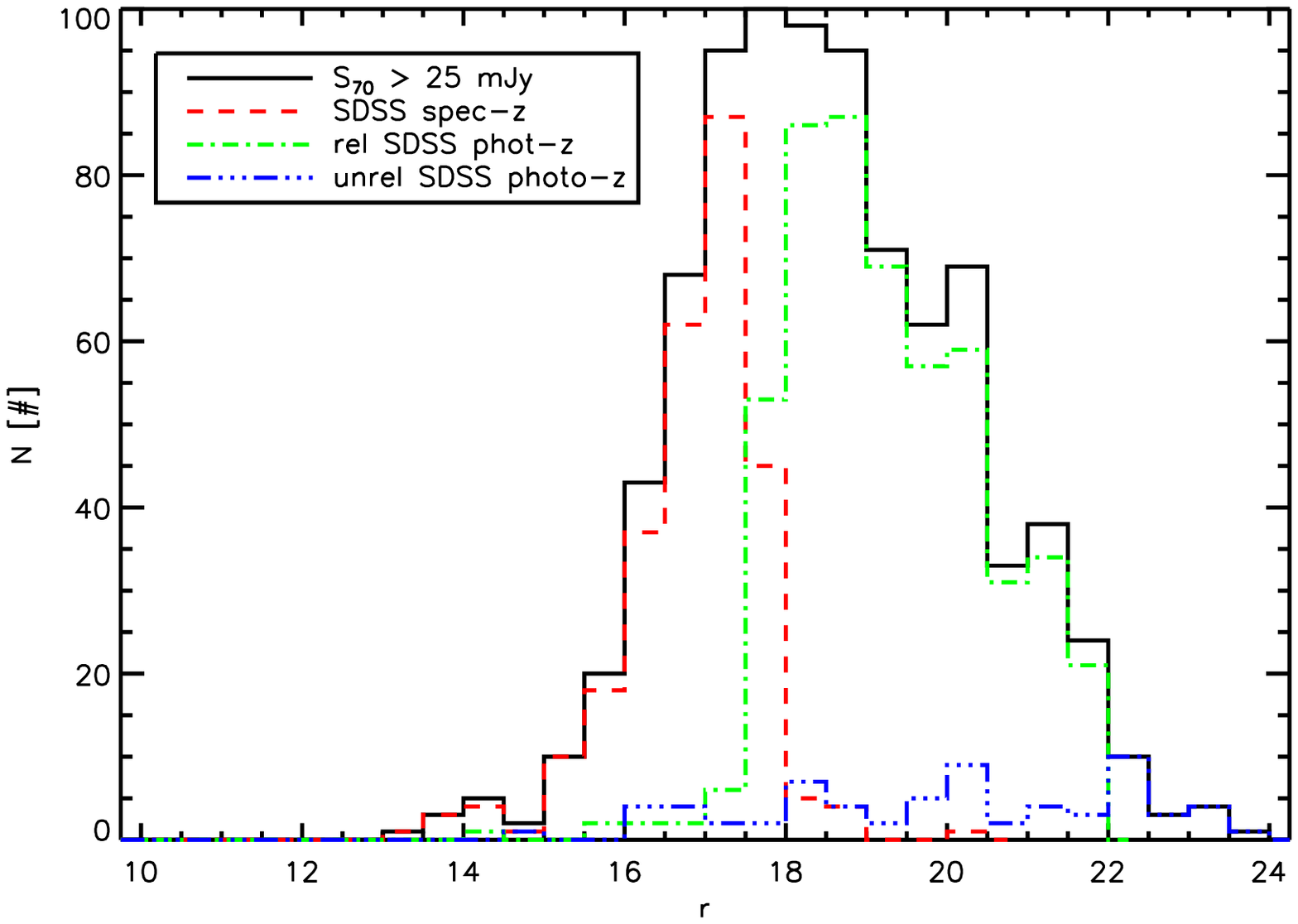}
\includegraphics[width=\linewidth]{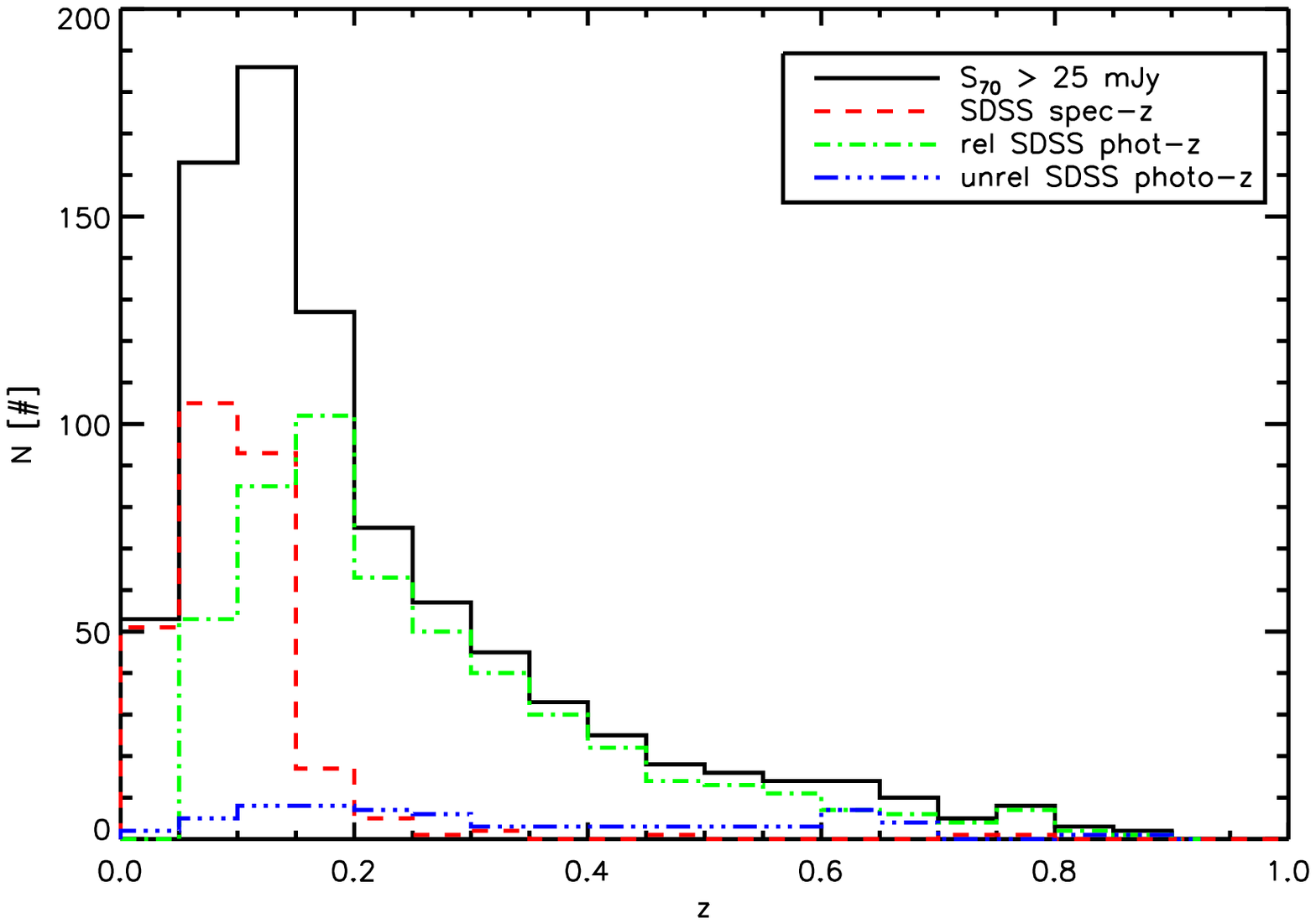}
\includegraphics[width=\linewidth]{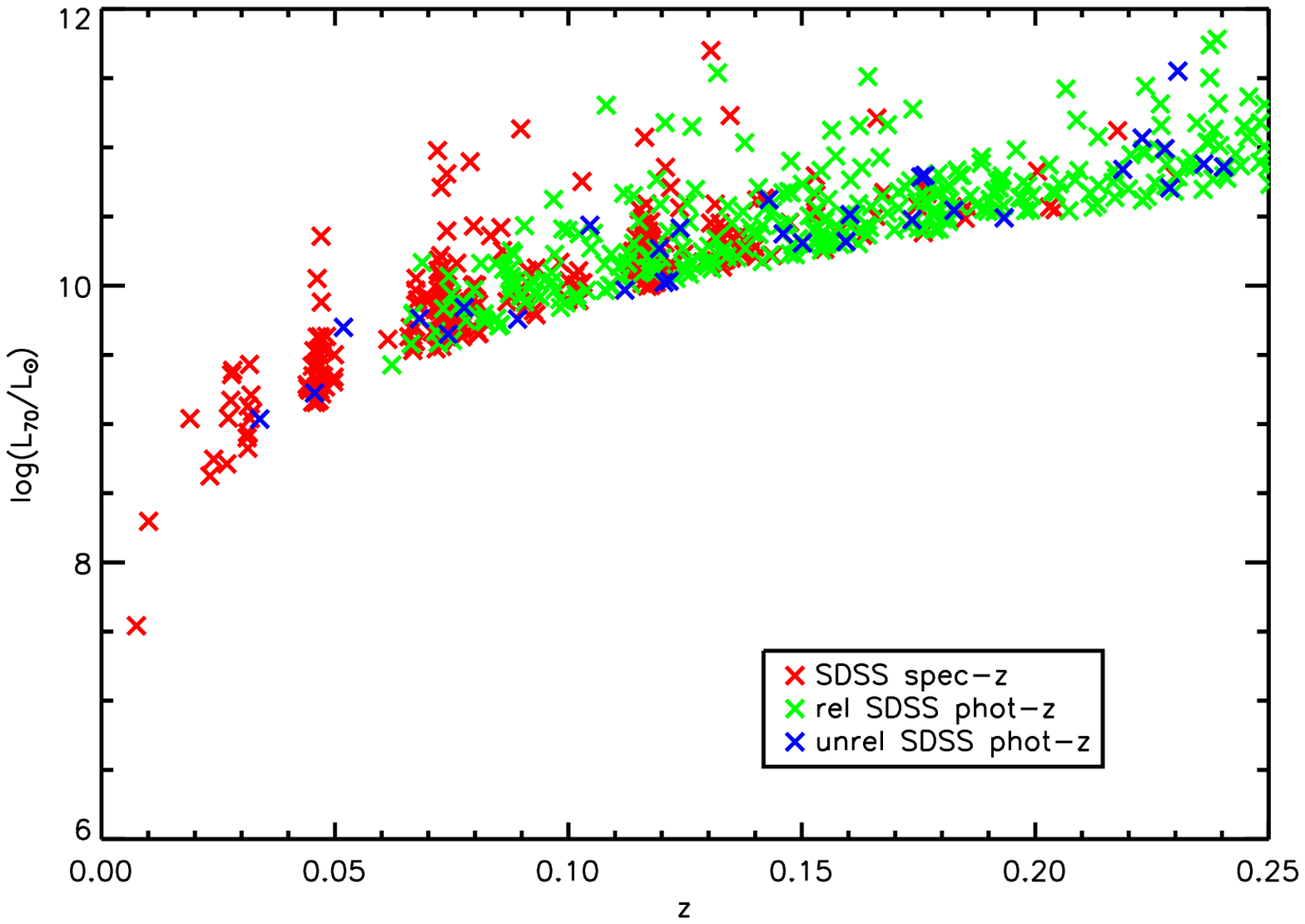}
\caption{The SWIRE-SDSS 70 micron sample. From the top : identification rate, magnitude, redshift and luminosity distribution.}
\label{fig:nmz70}
\end{figure}

The basic properties of the SWIRE-SDSS 70 micron sample, here selected with a flux cut of 25 mJy, which is where SWIRE completeness stars falling off, are illustrated in Figure~\ref{fig:nmz70}. For such relatively bright FIR sources, the optical identification rate (first panel) turns out to be very high, with most of the sources which failed to be reliably assigned either an optical counterpart or a redshift believed to reside at higher redshifts. Their magnitude distribution (second panel) sharply falls off well before the SDSS magnitude limit, guaranteeing an accurate magnitude (and thus redshift) measurement for most sources. The redshift distribution (third panel), although peaking at low redshifts, shows an extended tail, highlighting the potential of the SWIRE-SDSS FIR-selected database for investigating the higher-redshift universe and thus the evolution of the luminosity function from the local universe up to at least $z>0.5$. The fraction of spectroscopic redshifts decreases sharply at $z\sim0.15$ and $r\sim18$, as expected on the basis of SDSS spectroscopy selection criteria. The fraction of photometric redshifts deemed ''unreliable'' by the SDSS DR6 photometric redshift pipeline is limited throughout most of the redshift and magnitude range, increasing sharply only at $z>0.6$ and $r>22$. Finally, the luminosity distribution of local sources (fourth panel) spans more than three dex and thus allows to place robust constraints on the luminosity function.
\section{The SWIRE-SDSS FIR LLFs}
\begin{figure*}
\centering
\includegraphics[width=0.7\linewidth]{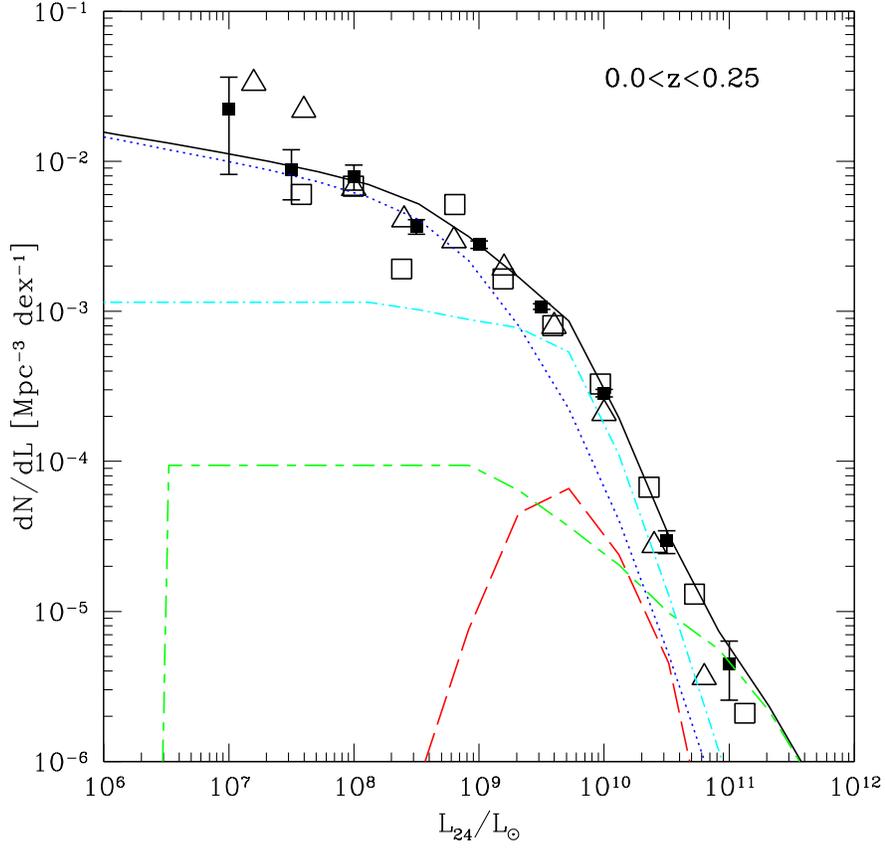}
\caption{SWIRE-SDSS MIPS LLF. 24 micron -
Filled Squares (This Work),
Empty Squares from Shupe et al. 1998,
Empty Triagles from Marleau et al. 2007.
Lines are models from Franceschini et al. (in prep). See text for details.}
\label{fig:24}
\end{figure*}

\begin{figure*}
\centering
\includegraphics[width=0.475\linewidth]{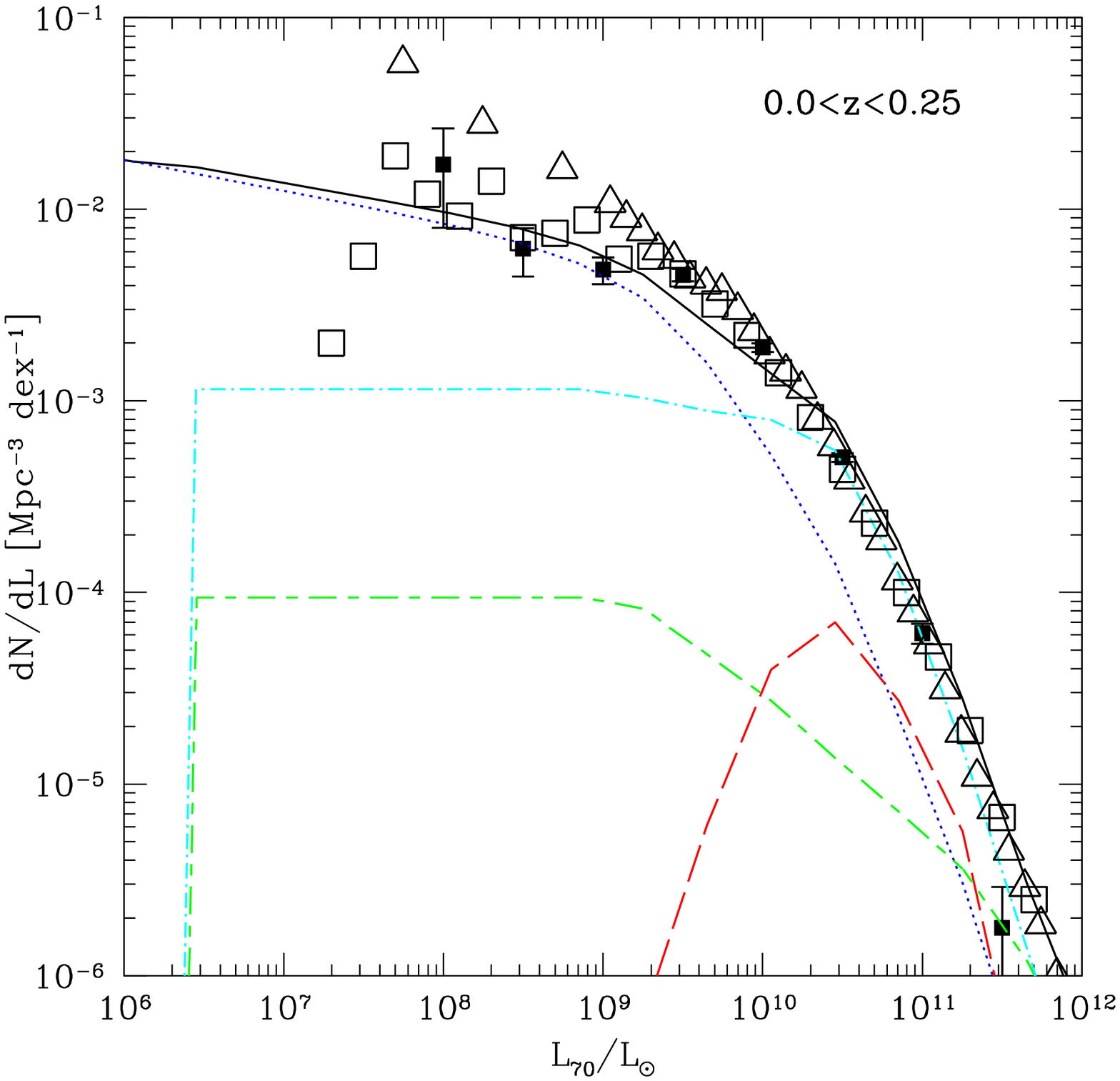}
\includegraphics[width=0.475\linewidth]{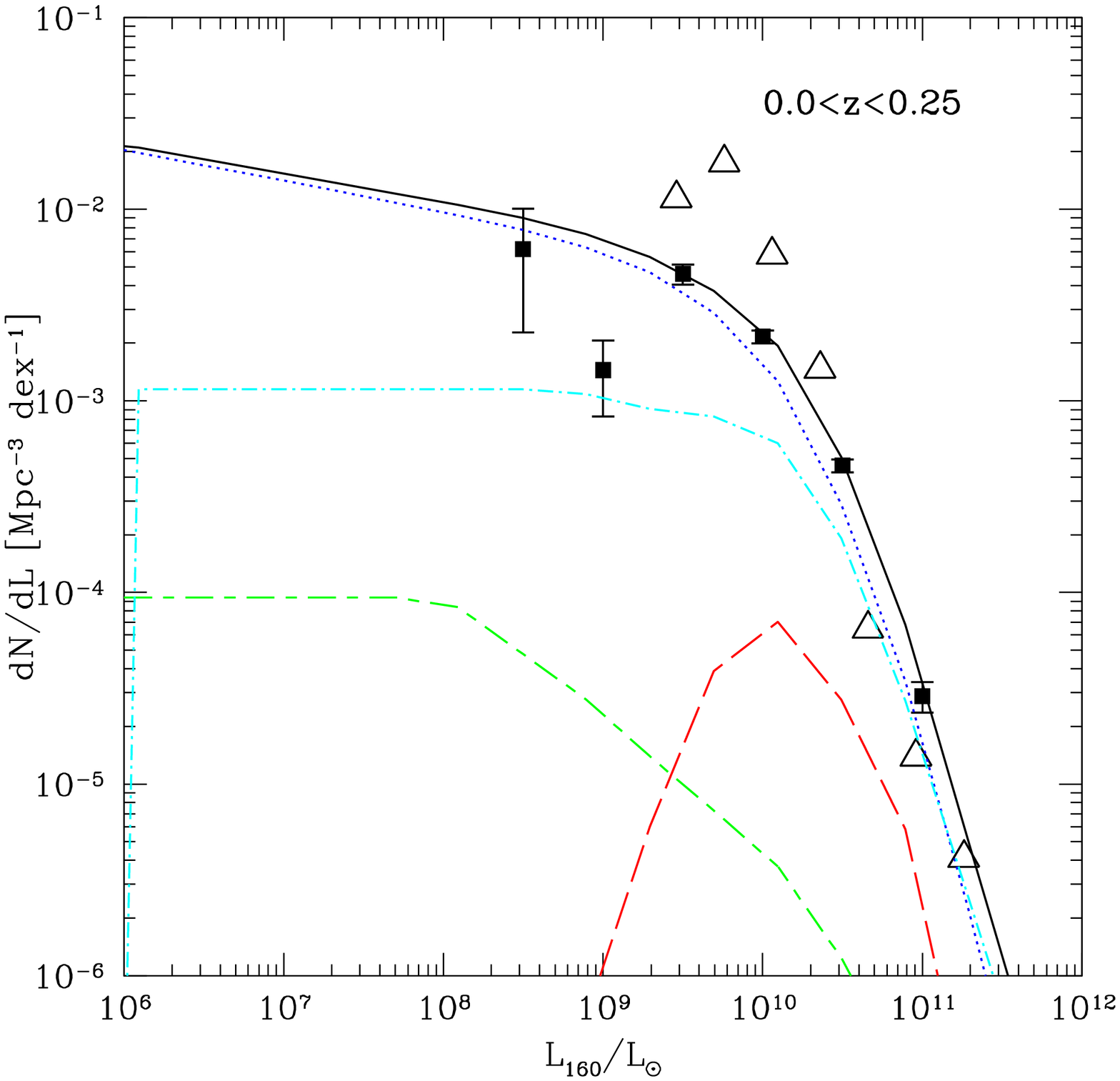}
\caption{SWIRE-SDSS MIPS LLF.
Left Panel - 70 micron - 
Filled Squares (This Work),
Empty Squares from Saunders et al. 1990,
Empty Triangles from Takeuchi et al. 2003.
Right Panel - 160 micron - 
Filled Squares (This Work),
Empty Triangles from Takeuchi et al. 2006.
Lines are models from Franceschini et al. (in prep). See text for details.}
\label{fig:70160}
\end{figure*}

Armed with the reliable and complete sample of bright FIR sources provided by the SWIRE-SDSS database, we computed the monochromatic luminosity functions over $0<z<0.25$ at 24, 70 and 160 micron applying the $1/V_{max}$ estimator to $S_{24,70,160} > 1,15,75$ mJy sources. A completeness correction based on simulations was applied to 70 and 160 micron sources fainter than 25 and 125 mJy respectively.
In Figures~\ref{fig:24}~and~\ref{fig:70160} we compare our results with previous work as well as with a backward evolution model for the evolution of FIR sources by Franceschini et al. (in prep). The black solid lines indicate the expected total LF, while the four contributing populations are respectively: slowly or non-evolving disk galaxies [blue dotted lines]; type-1 Active Galactic Nuclei evolving as shown by UV and X-ray selected quasars and Seyferts [green long-short dashed lines]; moderate-luminosity starburst galaxies with peak emission at $z\sim1$ and type-2 Active Galactic Nuclei [cyan dot-dashed lines]; ultra-luminous SCUBA-like starburst galaxies with peak evolution between $z=2$ and $z=4$ [red long dashed lines]. The agreement with the model is mostly very good, albeit within the uncertainties affecting the sample particularly at fainter luminosities. Additionally, our results overcome the large uncertainties characterizing some previous results. Note that errors in data points reported from previous work are not shown for clarity but are in most cases larger than ours, and that the full SWIRE-SDSS 70 micron sample will eventually include more than twice the sources used in this work.
\section{Harnessing the power of VO Tools}
This work provides further proof of the power of archival astronomy in the internet age by combining value-added data products by two large-area legacy survey projects in the infrared and optical wavelength range. Streamlining access and visualization of large datasets as well as data processing steps combining data acquired at different wavelength, VO tools have greatly contributed to enabling similar studies of an increased depth and breadth.

While the preliminary analysis of the SWIRE-SDSS FIR Local Luminosity Functions presented in this work relied on a set of ad hoc scripts mostly developed in IDL and SuperMongo, in order to refine and extend our results, we will be applying tools developed within the VO to streamline the workflow and make the results available to the community in a VO-compliant way.

An exploratory effort at reproducing the steps we followed in catalog inspection, visualization and cross-correlation using TopCat has not only made very rapid progress but also enabled much more thorough checks on various aspects of data quality thanks to TopCat built-in interactive plotting facilities. In the future we plan to exploit as much as possible the scripting capabilities provided by the VO Desktop and AstroGrid Python, and look forward to when these will support IDL scripting as well. This will arguably simplify to a large extent the implementation of a number of improvements, such as the extension to SWIRE EN1 and EN2 fields (thus almost doubling covered area), and the use of SDSS DR7 and UKIDSS DR5.

A reliable and extensible VO platform and the availability of high-quality well-documented datasets via online archives will be the main factor driving an increased adoption of VO tools by the community in the near future. An outstanding opportunity to get the community excited about the VO will be offered by the launch of Herschel, an ESA Cornerstone mission observing in the FIR \& SMM wavelength range and the first "great observatory" to be launched in the VO era. Herschel Data Processing tools will be tightly integrated with VO tools through a Plastic server, offering a seamless way to exchange data in tabular and image formats. This will ensure an optimal early science exploitation of Herschel data, and incidentally allow to detail and improve constraints on the FIR LLF estimates presented in this work.
%
%
%

%
%
\end{document}